\documentclass[%
 reprint,
superscriptaddress,
 amsmath,amssymb,
 aps,prl
]{revtex4-2}

\usepackage{graphicx}
\usepackage{dcolumn}
\usepackage{bm}
\usepackage{amsmath,amssymb}   
\usepackage{mathtools}         
\DeclareMathOperator{\sech}{sech}

\begin{document}

\preprint{APS/123-QED}

\title{Variable-mass dynamics of solitons in ferrimagnets}
\author{Pietro Diona}
\email{pietro.diona@sns.it}
\affiliation{Nanoscience, Scuola Normale Superiore, Piazza dei Cavalieri 7, Pisa, 56126, Italy}
\affiliation{Quantum Materials Theory, Italian Institute of Technology, Via Morego 30, Genoa, 16163, Italy}%
\author{Sergey Artyukhin}%
\email{sergey.artyukhin@gmail.com}
\affiliation{Quantum Materials Theory, Genoa, 16162, Italy}
\author{Luca Maranzana}
\email{luca.maranzana@iit.it}
\affiliation{Quantum Materials Theory, Italian Institute of Technology, Via Morego 30, Genoa, 16163, Italy}%
\affiliation{Physics, University of Genoa, Via Dodecaneso 33, Genoa, 16146, Italy}
\date{\today}
\begin{abstract}
Domain wall motion underpins emerging spintronic technologies, such as high-speed racetrack devices and THz logic, and accelerating walls quickly is a key challenge on the path to faster devices. Recent experimental advances introduced magnetic materials with non-uniform composition, allowing angular momentum compensation points and fast domain wall motion, although acceleration of domain walls in these materials remains poorly understood. Here, we show that spatial variation of exchange and anisotropy not only pushes the wall towards lower wall surface tension region, but also modifies its inertial mass, giving rise to a variable-mass relativistic dynamics. We find another force, originating from the magnon velocity gradient, that dominates as the wall velocity approaches the magnon speed. Our results identify domain walls in nonuniform magnets as a playground for relativistic physics with variable mass and limiting speed. 
\end{abstract}




\maketitle

\paragraph{Introduction ---}
Magnetic domains can serve as information carriers, hence the concept of racetrack memory proposed for the first time by Parkin et al. \cite{parkin2008magnetic}. While the early studies focused on ferromagnets, attention has progressively shifted toward materials that enable higher energy efficiency and faster domain wall motion, such as antiferromagnets \cite{baldrati2019mechanism, gomonay2016high, baltz2018antiferromagnetic, shiino2016antiferromagnetic} and ferrimagnets \cite{caretta2024domain, kim2022ferrimagnetic, vsteady1, diona2025observation}. In these materials, domain walls are relativistic sine-Gordon solitons, with the spin-wave group velocity playing the same role as the speed of light plays in the theory of special relativity \cite{einstein1905elektrodynamik, fogel1977dynamics, huang2015exact, diona2025observation, Maranzana25, Zvezdin79, caretta2020relativistic}. Beyond improving performance, real-world spintronic technologies require the ability to manipulate, synchronize, and confine domain walls. Initially, their manipulation was achieved via current-induced spin-transfer torques \cite{STT, brataas2012current}, and later a more efficient path using spin–orbit torques (SOT) was demonstrated \cite{manchon2019current, ryu2020current, ramaswamy2018recent, SOT1}. In parallel, non-uniform magnetic systems have been explored as a route to engineer spatial variations of fundamental material parameters, enabling local control of domain walls. By tailoring the spatial profile of the domain wall surface tension, it becomes possible to pin, accelerate or decelerate, and even synchronize domain walls \cite{diona2022, zhang2017spin}. Such spatial non-uniformities can be introduced either optically through non-uniform laser exposure \cite{riddiford2024grayscale} or electrically, via voltage-controlled magnetic anisotropy (VCMA) \cite{tan2019high}. These approaches were developed as alternatives to lithographically defined notches, which require high-precision fabrication to accurately control their depth \cite{diona2022, riente2017controlled, whang2018analytic}. As solitonic quasi-particles, domain walls possess inertia--—an ability to retain their momentum \cite{thomas2010dynamics}---enabling mechanisms for spin switching \cite{kimel2009inertia} and THz operation \cite{foggetti2025quantitative, neeraj2021inertial, tatara2020magnon, gomonay2016high, yan2011fast}. The lower mass of the domain wall allows faster acceleration\cite{Yan2010,yan2011fast}. Spatial variations in spin stiffness and anisotropy imply changes in the energy of the domain wall, allowing wall acceleration by the force defined by the gradient of wall energy \cite{li2020ultrafast,zhang2017spin}.

Here, we uncover two alternative acceleration mechanisms, one that relies on the momentum conservation in a variable mass system, and another on spatial variations of magnon velocity. We describe domain wall dynamics in non-uniform ferro- and  ferri- magnets and show that gradients of anisotropy or exchange also enable spatial engineering of the domain wall mass and local magnon velocity. As the wall travels along mass gradients, it accelerates not only due to forces but also due to its decreasing mass, analogous to the propulsion of a rocket. An additional force owing to the magnon speed gradient dominates in the relativistic limit, while all other forces tend to zero, as seen in  Fig.~\ref{fig:forces}. We derive an equation of motion describing these effects and corroborate it by micro-magnetic simulations.

\begin{figure}[b]
     \centering
         \includegraphics[width = \linewidth]{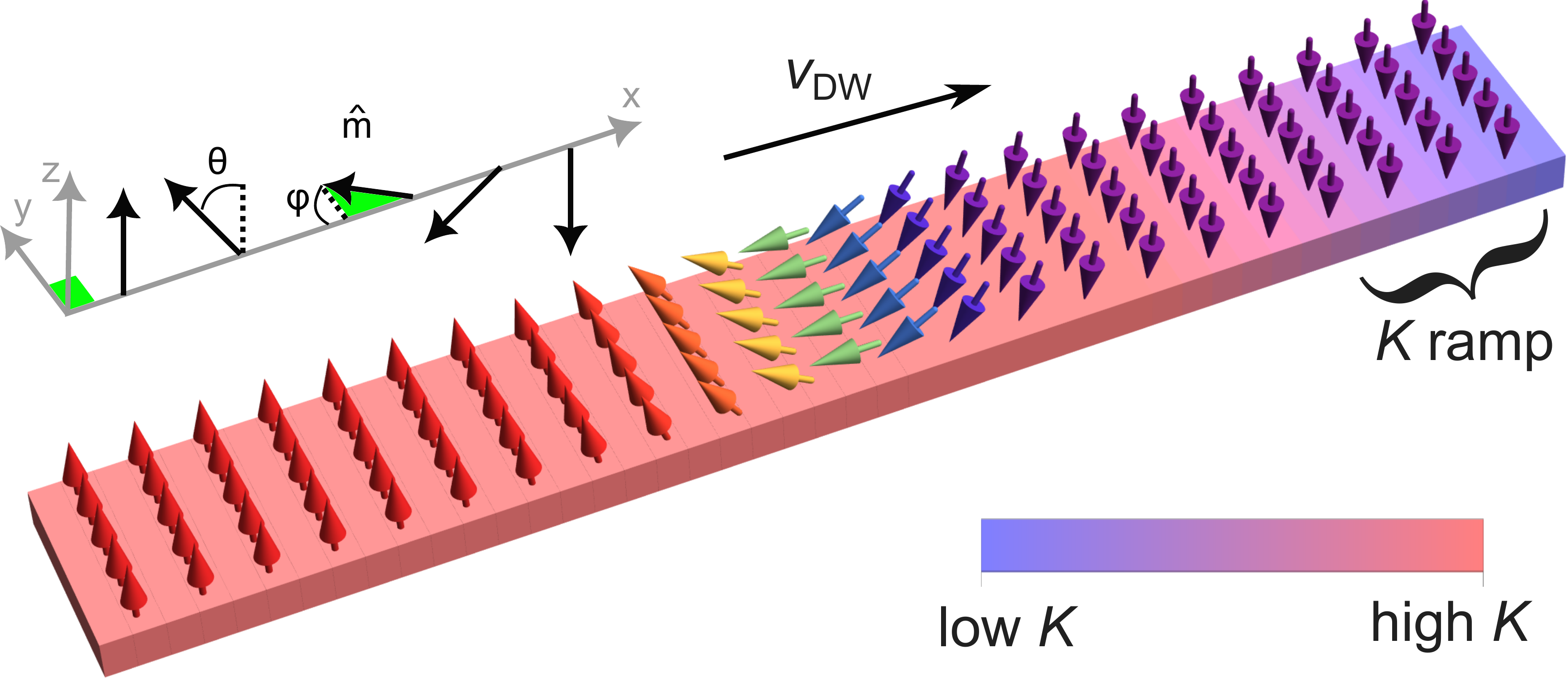}
    \caption{Schematic representation of a magnetic racetrack featuring a segment with spatially varying parameters.
The color gradient from blue to red denotes a gradual decrease of the magnetic anisotropy
$K$ (or increase of the exchange stiffness 
$A$) along the racetrack. A Néel-type domain wall separates a down-magnetized domain from an up-magnetized one. Areas in green are in $xy$ plane.}
       \label{geom} 
\end{figure}

\begin{figure}[t]
     \centering
         \includegraphics[width = 0.8\linewidth]{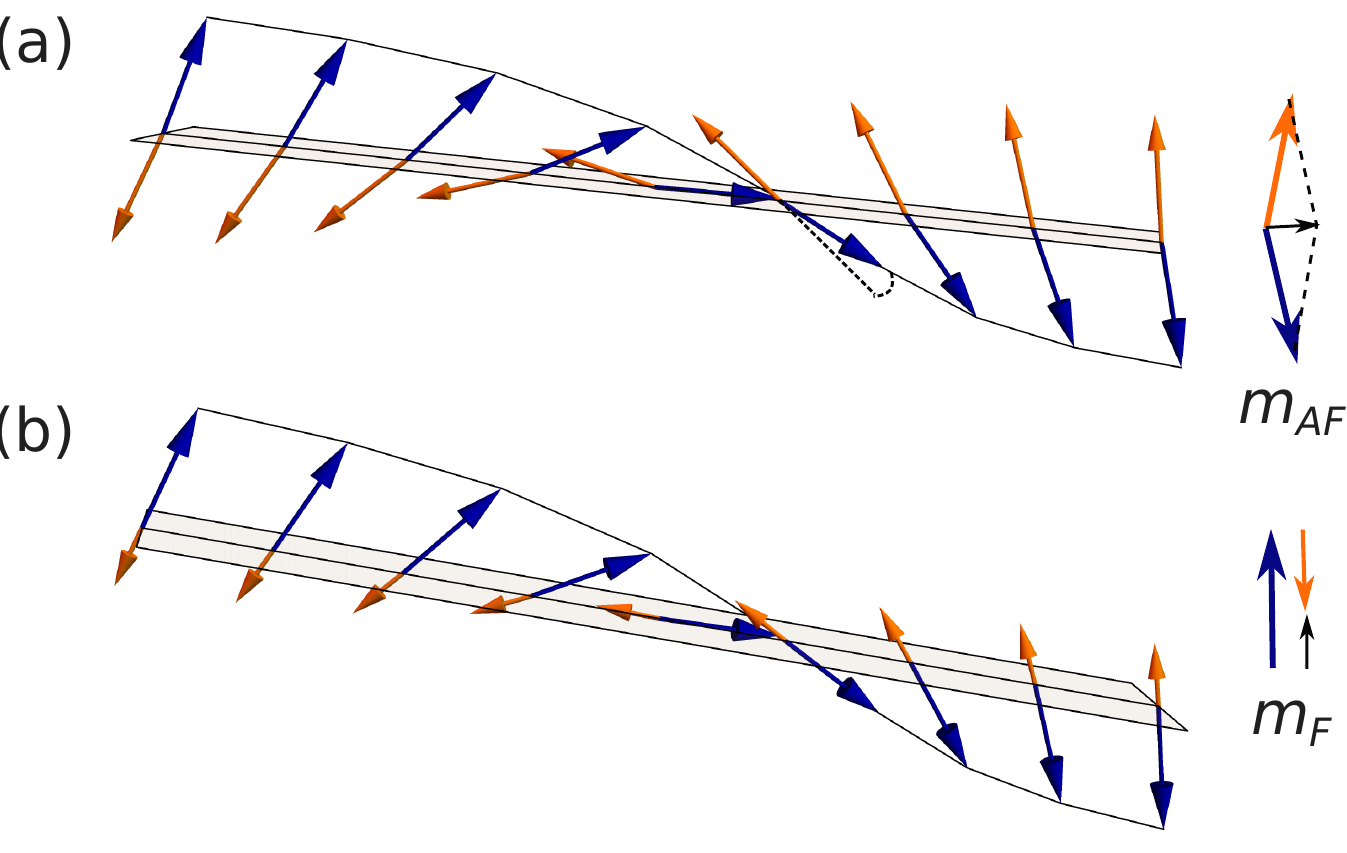}
    \caption{(a) The antiferromagnetic mass $m_{\mathrm{af}}$ originates from the net magnetic moment associated with spin canting between the sublattices, which generates a torque on the spins and is conjugate to the translational coordinate $q$ of the domain wall. Spins of AFM sublattices are in blue and yellow, and the canted magnetization is indicated by the black arrow. (b) The ferromagnetic mass $m_{\mathrm{f}}$ arises from the net magnetic moment due to uncompensated sublattices. This degree of freedom vanishes in the fully compensated limit.}
       \label{modes} 
\end{figure}

\paragraph{Ferromagnetic toy model ---}
We first illustrate the concept using a ferromagnetic toy model, whose free energy density is \cite{pap1, pap2, diona2022}: 
\begin{equation}
\label{hamiltonian}
\mathcal{F} =  A_{\mathrm{ex}}\left(\nabla\hat{m}\right)^2 + K \left(1-m_{z}^{2}\right) + \mathcal{K}_{d}m_{y}^{2} - M_s\vec{H}\cdot\hat{m},
\end{equation}
where \(\hat{m}\) is the local magnetization direction, coordinate-dependent $A_{ex}$ describes the ferromagnetic exchange interaction, $K$ denotes the easy-axis anisotropy, $\mathcal{K}_d$ is the transverse shape anisotropy that favors a Néel domain wall configuration, followed by the Zeeman term (see SI for details). The dynamics of the system is described by the following Lagrangian and Rayleigh densities: 
\begin{equation}
\label{lagrangian1}
    \mathcal{L} = J\bar{a}\left(\hat{m}\right)\cdot\dot{\hat{m}} - \mathcal{F}, \qquad 
    \mathcal{R} =  \frac{\alpha J}{2}\dot{\hat{m}}^2,
\end{equation}
where $\bar{a}(\hat{m})$ is the vector potential, $J$ is the density of angular momentum given by the ratio between the saturation magnetization and the gyromagnetic ratio, and $\alpha$ is the Gilbert damping. 

We study the dynamics of a domain wall with position $q$ and a small azimuthal angle $\phi$ (Fig.~\ref{geom}), assuming a time-dependent transverse shape anisotropy $\mathcal{K}_y(t)$. Applying the Euler-Lagrange-Rayleigh formalism, we obtain the equation of motion (see the SI for more details):
\begin{equation}
\label{mFerro}
    m_{\mathrm{f}}(t)\ddot{q} = -\frac{\alpha J}{\Delta}\dot{q} -\dot{m}_{\mathrm{f}}(t)\dot{q} - H_z M_{s},
\end{equation}
where $m_{\mathrm{f}}(t) = J^2/(2 \Delta \mathcal{K}_y(t))$ is the effective mass of the domain wall and $\Delta = \sqrt{A_{ex}/K}$ is its width. The azimuthal angle $\phi \propto m_{\mathrm{f}}(t)\dot{q}$ plays the role of momentum. Hence, as the ferromagnetic domain wall approaches the Bloch configuration, $\phi \rightarrow \pi/2$, its momentum increases. Conversely, a Néel domain wall ($\phi = 0$) has zero momentum. The term with $\dot{m}_f$ in Eq.~\ref{mFerro} describes an accelerating force if the domain wall loses its mass in time, analogously to a rocket gaining thrust through mass variation \cite{tsiolkovsky1954reactive}. The role of $\mathcal{K}_y(t)$ can be equivalently played by Dzyaloshinskii–Moriya interaction (DMI), or azimuthal Zeeman magnetic field. However, a time-varying Zeeman field, DMI or transverse shape anisotropy is experimentally challenging.

We therefore introduce a spatially non-uniform system, with space-dependent ferromagnetic exchange interaction $A_{ex}(x)$, easy-axis anisotropy $K(x)$, saturation magnetization $M_s(x)$, and transverse shape anisotropy $\mathcal{K}_y(x)$. We assume these parameters to vary smoothly in space over a length scale much larger than the domain wall width, ensuring that the domain wall shape remains undistorted. Under this assumption, the explicit $x$-dependence can be replaced by a dependence on the collective coordinate $q$, meaning that all parameters change as functions of the domain wall position. 
Consequently, the rocket term $-\dot{m}_{\mathrm{f}}(t)\dot{q}$ of Eq.~\ref{mFerro} becomes $- m'_{\mathrm{f}}(q)\dot{q}^2$, where
$m_{\mathrm{f}}(q)$ depends on time only implicitly through the domain wall position $q(t)$. This extra force is proportional to the wall mass gradient and the square of the velocity, hence becoming important at high velocities, which are limited in ferromagnets by the Walker breakdown.

\paragraph{Rocket effect in non-uniform ferrimagnets --- }
Ferrimagnets support ultrafast domain wall dynamics due to the suppression of Walker breakdown, making them promising candidates for next-generation magnetic technologies \cite{zhang2023ferrimagnets, foggetti2025quantitative, neeraj2021inertial}.
Under these conditions, extend the equation of motion of a ferromagnetic domain wall, Eq.~\ref{mFerro}, to the non-uniform ferrimagnetic case, showing that a domain wall can shed mass and accelerate toward the magnon velocity.

The Lagrangian and Rayleigh densities of a ferrimagnet with antiferromagnetic order parameter along a unit vector $\hat{n}$ read \cite{kim2017fast, li2020ultrafast, kim2014propulsion, kim2017self, pap2, caretta2020relativistic}: 
\begin{equation}
\label{nonUniformBerryFinal1}
\mathcal{L}\approx\frac{\rho (x)}{2}\dot{\hat{n}}^{2} + \delta_{s}(x)\bar{a}\left(\hat{n}\right)\cdot\dot{\hat{n}} - \mathcal{F}, \,\, \mathcal{R}= \frac{s_{\alpha}(x)}{2}\dot{\hat{n}}^{2},
\end{equation}
where \(\rho(x) = s_{T}^{2}(x) d^2/4A(x)\) parametrizes the antiferromagnetic inertia associated with the dynamics of \(\hat{n}\), $\bar{a}\left(\hat{n}\right)$ is the vector potential, \(s_{\mathrm{T}}(x) = s_{1}(x) + s_{2}(x)\) is the total spin density and \(\delta_s(x) = s_{1}(x) - s_{2}(x)\) is the net spin angular momentum density, \(d\) is the lattice constant, $A(x)$ describes the antiferromagnetic exchange interaction, and \(s_{\alpha} (x)\approx \alpha s_{\mathrm{T}}(x)\).
The free energy density of a non-uniform ferrimagnet is formally identical to Eq.~\ref{hamiltonian}, $\mathcal{F}(\hat{n},x) = A(x)\left(\nabla\hat{n}\right)^2 + K (x)\left(1-n_{z}^{2}\right) + \mathcal{K}_{d}n_{y}^{2} - M_s(x)\vec{H}\cdot\hat{n}$ \cite{oh2017coherent, li2020ultrafast, shiino2016antiferromagnetic, hubert1998magnetic}. 
Following the approach adopted in the previous section, we obtain the equation of motion of the following form (see the SI for more details):
\begin{equation}\label{eq:reldyn}
    \frac{d(m_{\mathrm{tot}}\gamma \dot{q})}{dt}=-\partial_q(\gamma^{-1}\sigma_0 )+2H_zM_s -\alpha s_\mathrm{T} \gamma \dot{q} \Delta_0^{-1},
\end{equation}
where $x$-dependence in $m_{\mathrm{tot}}$, $\gamma$, $\sigma_0$, $M_s$, and $s_T$ and $\Delta_0$ is replaced by the dependence on the wall position $q$, as discussed in SI; the relativistic contraction of the domain wall width is taken into account by the Lorentz factor $\gamma (q) = 1/\sqrt{1 - \left(\dot{q}/v_{g}(q)\right)^2}$ \cite{caretta2020relativistic, diona2025observation}, with $v_{g}(q) = \sqrt{2A (q)/(\rho_{\mathrm{f}}(q) + \rho (q))}$ the magnon speed and $\rho_{\mathrm{f}}(q) = \delta^2_s(q)/\mathcal{K}_d$ the ferromagnetic inertia. The total domain wall mass $m_{\mathrm{tot}}(q)$ is the sum of two contributions: the ferromagnetic mass $m_{\mathrm{f}}(q)$, discussed in the previous section, and the antiferromagnetic mass $m_{\mathrm{af}}(q) = \rho (q)/\Delta_0 (q)$, where  $\Delta_0(q) = \sqrt{A(q)/K(q)}$ is the non-contracted domain wall width. After taking the derivatives and restoring the $q$-dependence, the equation of motion takes the form:
\begin{eqnarray}
\label{finalEq1}
m_{\mathrm{tot}}(q)\ddot{q} &=& - \frac{\alpha s_{\mathrm{T}}(q)}{\Delta_0 (q)\gamma^2(q)}\dot{q} - \frac{m'_{\mathrm{tot}}(q)}{\gamma^2(q)}\dot{q}^2  -\frac{H_z^{\mathrm{eff}}(q)M_s (q)}{\gamma^3(q)} \nonumber\\&&-m_{\mathrm{tot}}(q)v'_{g}(q)v_{g}(q)\left(1-\frac{2\dot{q}^2}{v^2_{g}(q)}\right)\frac{\dot{q}^2}{v^2_g(q)},
\end{eqnarray}   
%
%
where the effective field, $H_z^{\mathrm{eff}}(q) = -2 H_z + \sigma'_0(q)/\left(M_s(q)\gamma(q)\right)$, incorporates both the applied magnetic field $H_z$ and the effect arising from the gradient of the domain wall energy, $\sigma_0(q) = \sqrt{4A(q)K(q)}$. The third term in Eq.~\ref{finalEq1} represents the rocket effect originating from the gradient of the total domain wall mass, while the force in the last term is induced by spatial variations in the magnon velocity. The latter term dominates for $\dot{q}\to v_g$, since the other terms tend to zero together with $\gamma^{-1}$ (see Fig.~\ref{fig:forces}), signifying that the relativistic particle cannot accelerate beyond the magnon speed. 
At $\dot{q} = v_g$, the equation reduces to $\ddot{q}=\dot{v}_g=v_g'v_g$, which adjusts the speed of the massless relativistic particle to the changing magnon speed, $v_g$. The force due to $v_g'$ term changes sign at $\dot{q}=v_g/\sqrt{2}$ due to the interplay of two opposite contributions, originating from terms with $\gamma$ and $\gamma^{-1}$ under derivatives in Eq.~\ref{eq:reldyn}. At the spin angular momentum compensation point $\delta_s = 0$, where the domain walls can approach the magnon speed \cite{diona2025observation}, the ferromagnetic mass vanishes while the antiferromagnetic one remains. The non-dissipative ($\alpha=0$) version of Eq.~\ref{finalEq1} can be obtained from a special relativity Lagrangian describing a variable-mass particle, $\mathcal{L}=2H_zM_s q-m_{\mathrm{tot}}v_g^2\sqrt{1-\dot{q}^2/v_g^2}$, where $m_{\mathrm{tot}}v_g^2=\sigma_0$ is the domain wall energy at rest.


\begin{figure}[t]
    \centering
    \includegraphics[width=0.95\linewidth]{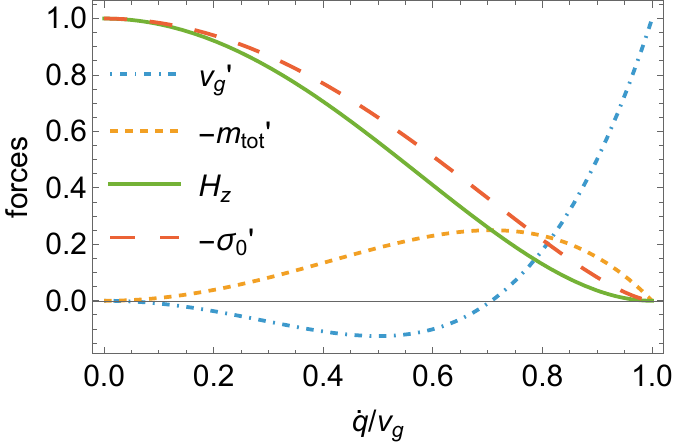}    \caption{\label{fig:forces}Velocity dependence of the force terms in Eq.~\ref{finalEq1}, $f(\dot{q}/v_g)$. The force due to an external magnetic field is marked by $H_z$, and due to the wall surface energy gradient by $-\sigma_0'$. The mechanisms introduced in this work, due to variable mass $-m_{\mathrm{tot}}'$ and due to the gradient of magnon velocity $v_g'$, have different functional dependency on the wall velocity. The latter mechanism dominates at high wall speed $\dot{q}\approx v_g$.}
\end{figure}

\begin{figure*}[t]
     \centering
         \includegraphics[width=.8\linewidth]{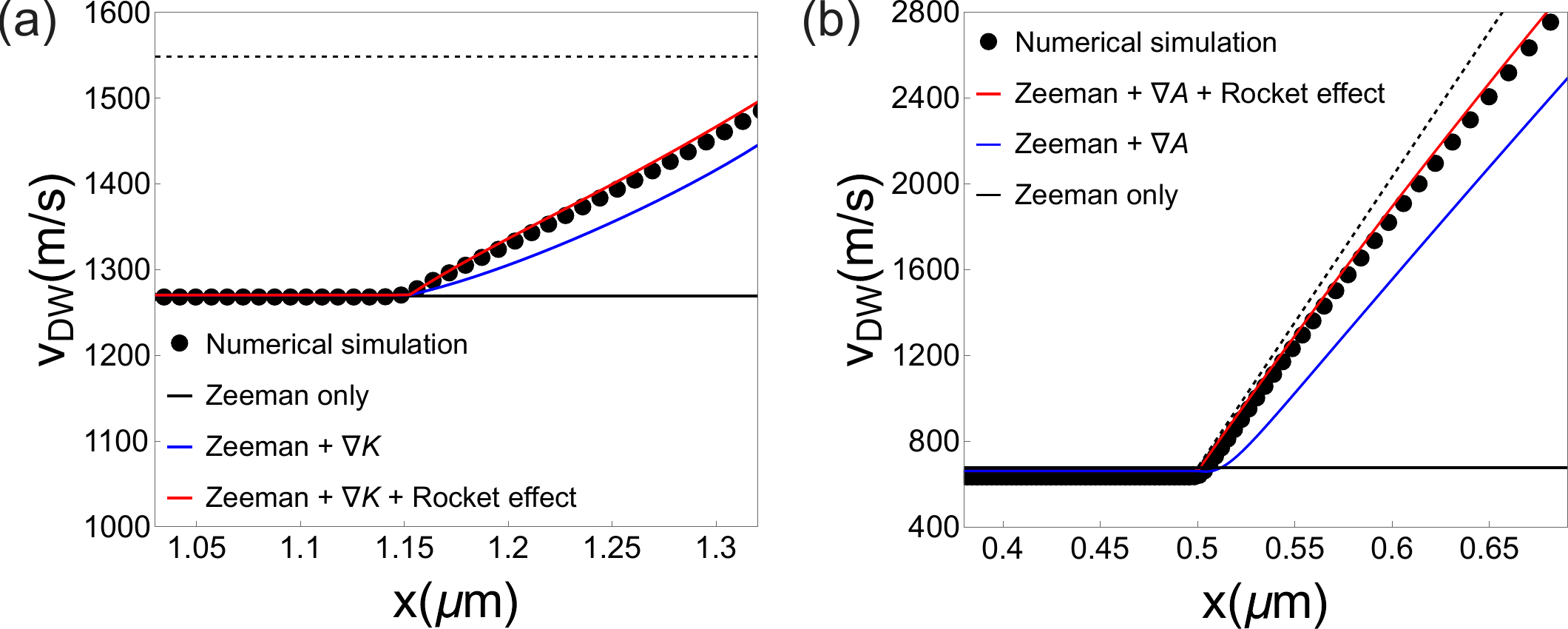}
         
    \caption{Rocket-like acceleration of a domain wall in a ferrimagnetic racetrack, where: (a) the anisotropy changes from $ 14\cdot 10^3\,\mathrm{J/m^3}$ to $ 5\cdot 10^3\,\mathrm{J/m^3}$ in $200 \, \mathrm{nm}$; (b) the antiferromagnetic exchange interaction varies from $ 1\,\mathrm{pJ/m}$ to $ 4\,\mathrm{pJ/m}$ in $200$ nm. The domain wall velocity is plotted as a function of its position along the racetrack. The numerical simulation (dotted black line) is compared with the full analytical model of Eq.~\ref{finalEq1} (red line), the same model neglecting the rocket effect (blue line), and considering only Zeeman field (black line). The dashed black lines indicate the magnon velocity. The initial velocity is acquired over the first (a) $1.15~\mu\mathrm{m}$ with a Zeeman field of 100 mT, and (b) $0.5~\mu\mathrm{m}$ with a Zeeman field of 150 mT. The reduction of the antiferromagnetic mass provides an additional “boost”, allowing the domain wall to approach the magnon velocity. The parameters of the simulation are reported in Table~S1, and refers to GdFeCo system \cite{diona2025observation}.}
       \label{sim}
\end{figure*}

The domain wall inertia originates from the magnetization component transverse to the wall plane, which serves as the canonical momentum conjugate to the wall position $q$. In the ferrimagnetic case, the antiferromagnetic contribution to the mass, $m_{\mathrm{af}}(q)$, arises from the dynamic canting between the sublattices that generates a small net magnetization associated with the Néel-order dynamics (see Fig.~\ref{modes}a). In contrast, the ferromagnetic contribution, $m_{\mathrm{f}}(q)$, stems from the equilibrium net magnetization due to sublattice imbalance \cite{kim2023mechanics, doring1948tragheit} (see Fig.~\ref{modes}b), which is coupled to the azimuthal (precessional) degree of freedom of the wall. The corresponding angle $\phi$ vanishes in the compensated limit, where this ferromagnetic inertia disappears. Thus, the total domain-wall mass reflects the dual dynamical character of ferrimagnets, combining inertial effects associated with both dynamic Néel-order canting and ferromagnetic precession.

\paragraph{Micromagnetic simulations --- }
To corroborate the rocket effect discussed in the previous section, we simulate a magnetic racetrack consisting of a uniform stretch used to accelerate the domain wall, and a 200‑nm‑long anisotropy/exchange ramp, schematically shown in Fig.~\ref{geom}. 
The domain wall is first accelerated by an external magnetic field toward the ramp region, where the anisotropy (the antiferromagnetic exchange interaction) is gradually reduced (increased). We set the operational regime of the ferrimagnet at the spin angular momentum compensation point, where Walker breakdown is suppressed and the domain wall speed is maximized \cite{diona2025observation, kim2017fast}. As a consequence, the ferromagnetic mass results to be negligible with respect to the antiferromagnetic one. As the domain wall traverses the ramp, the corresponding decrease in its antiferromagnetic mass gives rise to a ``rocket boost" (see Fig.~\ref{sim}). This effect drives the domain wall even closer to the magnon speed, highlighting a new mechanism for ultrafast domain wall acceleration. 
Due to a higher power of $A$ in $m_{\mathrm{af}} \propto K^{1/2}A^{-3/2}$, giving $2\delta m_{\mathrm{af}}/m_{\mathrm{af}}=(\delta K/K-3\delta A/A)$, the rocket effect is more pronounced due to a relative change in the exchange interaction (Fig.~\ref{sim}b) than the same relative change of the anisotropy (Fig.~\ref{sim}a). In the relativistic limit $\dot{q}\approx v_g$, the last term in Eq.~\ref{finalEq1} dominates, and is essential to obtain an agreement between the simulations and theory at high $\dot{q}$.

The simulation uses realistic material parameters reported in Table~S1, which are consistent with the experimental results in \cite{diona2025observation, sala2023deterministic}. Anisotropy values of approximately $10^4$~J/m$^3$ have been measured in amorphous ferrimagnetic systems \cite{diona2025observation, sala2022asynchronous, sala2022ferrimagnetic}, and spatial variations of up to $\pm 20-40\%$ can be induced \cite{maruyama2009large, diona2022, zhang2017spin, kato2018giant, diona2025observation}. Antiferromagnetic exchange interaction is varied between 1~pJ/m and 4~pJ/m, in agreement with an experimental window close to the spin angular momentum compensation point \cite{diona2025observation, brunsch1977perpendicular, katayama1978annealing, nishihara1979effects, joo2021magnetic}. The effect of the transverse shape anisotropy $\mathcal{K}_d$ can be replaced by an azimuthal magnetic field or DMI. The agreement between theory and simulations with realistic parameters, shown in Fig.~\ref{sim}, confirms the validity of the rocket effect in non-uniform ferrimagnetic systems.

\paragraph{Discussion --- }\label{sec12}

A promising platform for the experimental observation of the rocket effect is offered by amorphous rare-earth -- transition metal (RE-TM) ferrimagnets. By tuning their stoichiometry, one can adjust the anisotropy, the net magnetization, and the net spin angular momentum, which are the key ingredients for achieving domain wall velocities of several km/s \cite{caretta2018fast, cai2020ultrafast, kim2017fast, sala2022asynchronous}. In particular, RE-TM ferrimagnets combine weak perpendicular anisotropy, a large anomalous Hall response, and a low Gilbert damping parameter \cite{hansen1991magnetic, kim2019low, bainsla2022ultrathin, kato2008compositional}. These properties make this alloy an excellent candidate not only for accessing relativistic domain wall dynamics \cite{diona2025observation}, but also for observing the rocket effect. 
A practical route to engineer the required spatial inhomogeneity is through non-uniform laser exposure or VCMA, which can generate nanoscale anisotropy gradients. A domain wall can be initially accelerated by either a Zeeman field or SOT \cite{diona2025observation}; once it encounters the anisotropy gradient, it begins to lose mass, thereby accelerating further toward the magnon velocity (as shown in Fig.~\ref{sim}a).
An alternative strategy relies on optical patterning. Selective exposure of RE-TM nanowire can induce a spatial gradient in the relative concentrations of the two sublattices, thereby producing corresponding gradients in the magnetic parameters \cite{riddiford2024grayscale}. As demonstrated in \cite{diona2025observation}, the parameter most sensitive to the RE–TM composition is the antiferromagnetic exchange, whose magnitude can differ even for stoichiometric variations as small as $1\%$ \cite{raasch1994exchange, kato2008compositional, joo2021magnetic}. It is therefore realistic to engineer a sample in which the exchange stiffness varies from about 1 pJ/m to 4 pJ/m near the spin angular momentum compensation point \cite{diona2025observation, riddiford2024grayscale}. When domain walls in such a system are driven via SOT or Zeeman fields from the low-exchange region to the higher one,  antiferromagnetic mass is decreased, as shown in Fig.~\ref{sim}(b).
Although spatial composition variations also imply a spatial dependence of the sublattice spin densities $s_{\mathrm{1}}$ and $s_{\mathrm{2}}$ in a realistic system, this effect is neglected in a window that crosses the spin angular momentum compensation point. In fact, the latter are fixed in the simulation shown in Fig.~\ref{sim}(b), since in that window the ferromagnetic mass $m_{\mathrm{f}}$ is negligible compared to the antiferromagnetic one $m_{\mathrm{af}}$.

\paragraph{Conclusions --- }
Two acceleration mechanisms in magnetic domain wall dynamics are identified: one originates from spatial variations of domain wall mass and the other from gradients of magnon velocity. The generalized theory of domain wall motion for ferrimagnets with non-uniform exchange and anisotropy takes the form of the dynamics of a relativistic variable-mass particle traveling in the medium with limiting speed gradient. When the Walker breakdown is suppressed by a hard-axis anisotropy, in-plane field, or tuning the system toward the spin angular momentum compensation point, exchange and anisotropy gradients act as a rocket effect, a relativistic inertial boost, capable of pushing domain walls toward the velocity ceiling. The quantitative agreement between our analytical predictions and numerical simulations confirms the robustness of this effect. The experimental landscape is particularly promising. Amorphous ferrimagnets, such as GdFeCo, naturally offer the tunability needed to engineer steep nanoscale variations in exchange and anisotropy—through voltage control, optical patterning, or compositional gradients, providing a realistic route to observe mass shedding and ultrafast domain wall acceleration. 
Our findings introduce mass control as a new degree of freedom in ferrimagnetic soliton physics, expanding the design space for ultrafast magnetic devices. Magnetic domain walls are identified as a promising playground for relativistic variable-mass physics. Beyond its conceptual significance, the rocket effect suggests new strategies for high-velocity racetrack memories, THz emitters and ultrafast spintronic technologies.
\bibliography{sn-bibliography}


\section*{Author contributions}
S.A. conceived the project. P.D. and L.M. wrote the first draft. All authors edited the manuscript and prepared the figures. 

\section*{Competing interests}
The authors declare no conflict of interest.

\clearpage

\renewcommand{\theequation}{S\arabic{equation}}
\renewcommand{\thetable}{S\arabic{table}}
\renewcommand{\thefigure}{S\arabic{figure}}
\setcounter{equation}{0}
\setcounter{table}{0}
\setcounter{figure}{0}
\setcounter{section}{0}
\widetext
\section*{Supplementary information}
\section{Domain wall motion in non-uniform ferromagnets}

\begin{figure}
     \centering
         \includegraphics[width = 0.6\textwidth]{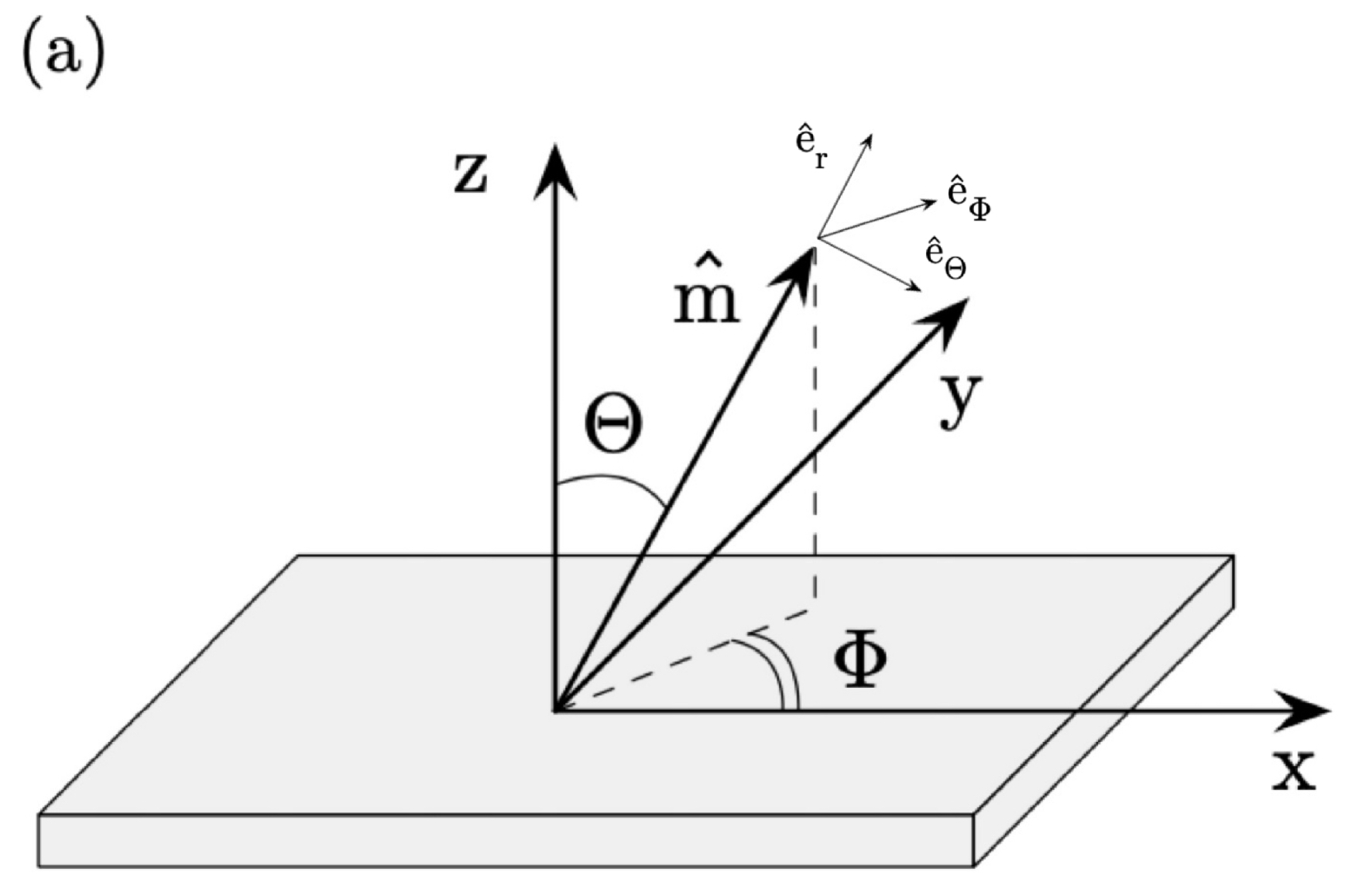}
         \includegraphics[width = 0.6\textwidth]{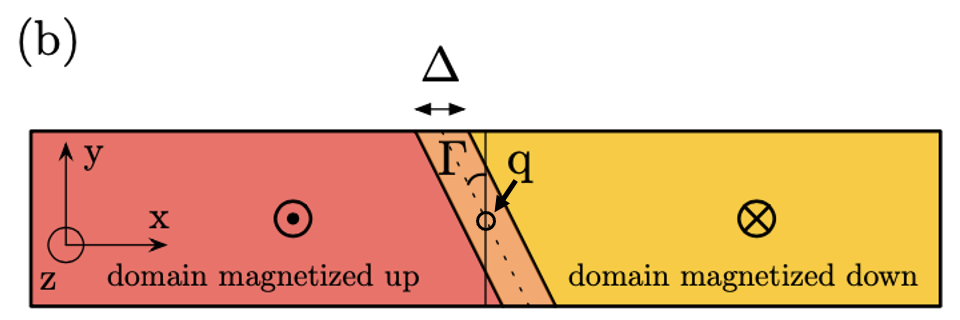}
    \caption{(a) Reference system; (b) Schematic representation of a magnetic domain wall. $\Delta$ is the domain wall width, $\Gamma$ is the tilting angle of the domain wall neglected by our analytical model, $q$ is the center of the domain wall \cite{diona2022}.}
       \label{refsystem} 
\end{figure}

The total Lagrangian density function for a topological soliton in a one dimensional non-uniform ferromagnetic system is \cite{pap1, pap2, diona2022}: 
\begin{equation}
\label{lagrangian}
    \mathcal{L} = J(x)\phi\dot{\theta}\sin{\theta} - \mathcal{E}\left(\hat{m},x\right).
\end{equation}
\(\mathcal{E}\left(\hat{m},x\right)\) is the total free energy density of the system, \(\phi\) is the precessional angle of the domain wall, \(\theta\) constitutes the out of plane angle, and \(J(x)\) is the non-uniform angular momentum density, given by the ratio between the saturation magnetization and the gyromagnetic ratio. The total free energy density is equal to \cite{pap1, pap2, diona2022}:
\begin{equation}
\label{hamiltonian1}
\mathcal{E}\left(\hat{m},x\right) =  A_{ex}(x)\nabla\hat{m}^2 + K(x) \left(1-m_{\mathrm{z}}^{2}\right) + \mathcal{K}_{y}(x)m_{\mathrm{y}}^{2} - M_s(x)\vec{H}\cdot\hat{m},
\end{equation}
where \(\hat{m}\) is the local magnetization direction, \(A_{ex}(x)\) describes the ferromagnetic exchange interaction, $K(x)$ denotes the easy-axis anisotropy density, $\mathcal{K}_y(x)$ is the transverse shape anisotropy density that favors a Néel domain wall state, and \(M_s(x)\vec{H}\cdot \hat{m}\) is the Zeeman term.
The system is non-conservative because of the damping effect, which can be taken into account with the Rayleigh dissipation function \cite{pap1, pap2, diona2022}: 
\begin{equation}
\label{rayleigh1}
    \mathcal{R} = \alpha \frac{J(x)}{2}\dot{\vec{m}}^2,
\end{equation}
where \(\alpha\) is the damping factor of the system. Euler-Lagrange-Rayleigh minimization of Eq.~\ref{lagrangian} and Eq.~\ref{rayleigh1} gives the the following ansatz solution for the polar angle $\theta$ \cite{pap1}: 
\begin{equation}
\label{ansatz1}
    \theta(x,t) = 2\arctan{\left[\exp{\left[\frac{x-q(t)}{\Delta(x)}\right]}\right]},
\end{equation}
where \(q(t)\) is the domain wall center, and \(\Delta(x) = \sqrt{\frac{A_{ex}(x)}{K(x)}}\) is the domain wall width. 
Plugging-in the ansatz (Eq.~\ref{ansatz1}) into the Lagrangian (Eq.~\ref{lagrangian}) and Rayleigh (Eq.~\ref{rayleigh1}) density functionals and moving into a spherical coordinates reference system leads to (see Fig.~\ref{refsystem} for more details): 
\begin{equation}
\label{list2}
\begin{cases}
    \mathcal{E}_A(x,t) = A_{ex}(x)\mathcal{D}^{'2}(x,t)\sech^2{\left[\mathcal{D}(x,t)\right]},\\
    \mathcal{E}_K(x,t) = K(x)\sech^2{\left[\mathcal{D}(x,t)\right]},\\
    \mathcal{E}_{\mathcal{K}_y}(x,t) = \mathcal{K}_y (x) \sech^2{\left[\mathcal{D}(x,t)\right]},\\
    \mathcal{E}_{Z}(x,t) = -\mu_{0}M_{s}(x)H_z\tanh{\left[\mathcal{D}(x,t)\right]},\\
    \mathcal{G}(x,t) = -J(x)\phi(t)\sech^2{\left[\mathcal{D}(x,t)\right]}\frac{\dot{q}(t)}{\Delta(x)},\\
    \mathcal{R}(x,t) = \alpha\frac{J(x)}{2}\sech^2{\left[\mathcal{D}(x,t)\right]}\left[\frac{\dot{q}^2(t)}{\Delta^{2}(x)} + \dot{\phi}^2\right],\\
\end{cases}
\end{equation}
where $M_s(x)$ is the non-uniform saturation magnetization and:
\begin{equation}
\label{definition}
    \begin{cases}
        \mathcal{D}(x,t) = \frac{x-q(t)}{\Delta(x)},\\
        \mathcal{D}^{'}(x,t) = \frac{\left[[q(t)-x]\Delta^{'}(x) + \Delta(x)\right]}{\Delta^{2}(x)}.
    \end{cases}
\end{equation}

\subsection{Two collective coordinates model}
\label{twocoordinate}
For the derivation of the collective coordinate model, we plug-in the ansatz (Eq.~\ref{ansatz1}) into the Lagrangian (Eq.~\ref{lagrangian}) and Rayleigh (Eq.~\ref{rayleigh1}) density, then we integrate in space. Finally by Euler-Lagrange-Rayleigh minimization, we get \cite{pap1, pap2,mougin2007domain}:
\begin{equation}
\label{2coord}
    \begin{cases}
        \frac{\alpha I_{1}}{\gamma}\dot{q} + \frac{I_{2}}{\gamma}\dot{\phi} = -\frac{\partial I_{3}}{\partial q} -\frac{\partial I_{4}}{\partial q} - \cos^2{\phi}\frac{\partial I_{5}}{\partial q} + \mu_{0}H_{z}\frac{\partial I_{6}}{\partial q},\\
        -\frac{I_{2}}{\gamma}\dot{q} + \frac{\alpha I_{7}}{\gamma}\dot{\phi} + I_{6}\sin{2\phi} = 0,
    \end{cases}
\end{equation}
where: 
\begin{equation}
\label{integral2coord}
    \begin{cases}
        I_{1} = \int_{l} \frac{M_s(x)\sech^2{\left[\mathcal{D}(x,t)\right]}}{\Delta^{2}(x)} dx,\\
        I_{2} = \int_{l} \frac{M_s(x)\sech^2{\left[\mathcal{D}(x,t)\right]}}{\Delta(x)} dx,\\
        I_{3} = \int_{l} A_{ex}(x)\mathcal{D}^{'2}(x,t)\sech^2{\left[\mathcal{D}(x,t)\right]} dx,\\
        I_{4} = \int_{l} K(x)\sech^2{\left[\mathcal{D}(x,t)\right]} dx,\\
        I_{5} = \int_{l} \mathcal{K}_{y}(x)\sech^2{\left[\mathcal{D}(x,t)\right]} dx,\\
        I_{6} = \int_{l} M_{s}(x)\tanh{\left[\mathcal{D}(x,t)\right]} dx,\\
        I_{7} = \int_{l} M_{s}(x)\sech^2{\left[\mathcal{D}(x,t)\right]} dx.\\  
    \end{cases}
\end{equation}
The integrals in Eq.~\ref{integral2coord} can be simplified in the hypothesis of slow variation of the parameters in space; i.e. they must not affect the domain wall ansatz. Therefore, the integrating functions in Eq.~\ref{integral2coord} results to be exponentially localized in a region where the parameters can be considered constant. It means that the fundamental parameters vary as a function of the domain wall center $q$ as follows: 
\begin{equation}
    \begin{cases}
        M_{s}(x) \approx M_{s}(q),\\
        A_{ex}(x) \approx A_{ex}(q),\\
        K(x) \approx K(q),\\
        \mathcal{K}_y(x) \approx \mathcal{K}_y(q),\\
        \Delta(x) \approx \Delta(q).\\
    \end{cases}
\end{equation}
Eq.~\ref{integral2coord}, in the hypothesis of an infinite one dimensional device simplifies as: 
\begin{equation}
\label{integralq}
    \begin{cases}
    I_{1} = \int_{l} \frac{M_{s}(q)\sech^2{\left[\frac{x-q(t)}{\Delta(q)}\right]}}{\Delta^{2}(q)} dx = \frac{M_{s}(q)}{\Delta^{2}(q)}\int_{-\infty}^{+\infty}\sech^2{\left[\frac{x-q(t)}{\Delta(q)}\right]}dx = 2\frac{M_{s}(q)}{\Delta(q)},\\
    I_{2} = \int_{l} \frac{M_s(x)\sech^2{\left[\mathcal{D}(x,t)\right]}}{\Delta(x)} dx = \frac{M_{s}(q)}{\Delta(q)}\int_{-\infty}^{+\infty}\sech^2{\left[\frac{x-q(t)}{\Delta(q)}\right]}dx = 2M_{s}(q),\\
    I_{3} = \int_{l} A_{ex}(x)\mathcal{D}^{'2}(x,t)\sech^2{\left[\mathcal{D}(x,t)\right]} dx = \frac{A_{ex}(q)}{\Delta^{2}(q)}\int_{-\infty}^{+\infty} \sech^2{\left[\frac{x-q(t)}{\Delta(q)}\right]} dx = 2\frac{A_{ex}(q)}{\Delta(q)},\\
    I_{4} = \int_{l} K(x)\sech^2{\left[\mathcal{D}(x,t)\right]} dx = K(q)\int_{-\infty}^{+\infty}\sech^2{\left[\frac{x-q(t)}{\Delta(q)}\right]}dx = 2K(q)\Delta(q),\\
    I_{5} = \int_{l} \mathcal{K}_y(x)\sech^2{\left[\mathcal{D}(x,t)\right]} dx = \mathcal{K}_y(q)\int_{-\infty}^{+\infty}\sech^2{\left[\frac{x-q(t)}{\Delta(q)}\right]}dx = 2\mathcal{K}_y(q)\Delta(q),\\
    I_{6} = \int_{l} M_{s}(x)\tanh{\left[\mathcal{D}(x,t)\right]} dx = M_{s}(q)\int_{-\infty}^{+\infty}\tanh{\left[\frac{x-q(t)}{\Delta(q)}\right]}dx = 2 q M_{s}(q),\\
    I_{7} = \int_{l} M_{s}(x)\sech^2{\left[\mathcal{D}(x,t)\right]} dx = M_{s}(q)\int_{-\infty}^{+\infty}\sech^2{\left[\frac{x-q(t)}{\Delta(q)}\right]}dx = 2M_{s}(q)\Delta(q).\\   
    \end{cases}
\end{equation}
Substituting Eq.~\ref{integralq} into Eq.~\ref{2coord}:
\begin{equation}
\label{2coordq}
    \begin{cases}
        \frac{\alpha J(q)}{\Delta(q)}\dot{q} + J(q)\dot{\phi} +
        H^{\mathrm{eff}}_z(q)M_s (q)= 0,\\[2ex]
        -\frac{J(q)}{\Delta(q)}\dot{q} + \alpha J(q)\dot{\phi} + \mathcal{K}_y(q)\sin{\left(2\phi\right)} = 0,
    \end{cases}
\end{equation}
where $H^{\mathrm{eff}}_z(q) = -\frac{A_{ex}^{'}(q)}{\Delta(q)M_s (q)} - \frac{\Delta (q) K^{'}(q)}{M_s (q)}$ is the effective magnetic field which arises from the gradient of $A_{ex}(q)$ and $K(q)$.
In the hypothesis of small $\phi$ angle, the second equation of system~\ref{2coordq} reduces to:
\begin{equation}
    \label{PhiD0}
\phi = \frac{J(q)}{2\Delta(q) \mathcal{K}_y (q)}\dot{q},
\end{equation}
from which:
\begin{equation}
    \label{phiDotD0}
    \dot{\phi} = \frac{J(q)}{2\Delta(q)\mathcal{K}_y(q)}\ddot{q} + \frac{\dot{\mathcal{K}}_y(q)}{\mathcal{K}_y(q)}\phi\dot{q}.
\end{equation}
Replacing both Eq.~\ref{PhiD0} and Eq.~\ref{phiDotD0} into the first equation of system~\ref{2coordq} and neglecting the Zeeman driving term, leads to the ferromagnetic domain wall equation of motion:
\begin{equation}
\label{ferroRocket}
    m_{\mathrm{f}}(q)\ddot{q} + \frac{\partial {m}_{\mathrm{f}}(q)}{\partial q}\dot{q}^2 +\frac{\alpha M_s}{\gamma\Delta}\dot{q} +  H^{\mathrm{eff}}_z(q)M_s (q) = 0,
\end{equation}
where $m_{\mathrm{f}} = \frac{J^2(q)}{2\Delta(q)\mathcal{K}_y (q)}$. The second term in Eq.~\ref{ferroRocket} shows that the non-uniform ferromagnetic domain wall can lose its mass during its motion. This physical effect can act as an effective acceleration/deceleration on the domain wall itself, exactly as it happens when a rocket is launched into space.

\subsection{Time dependent parameters}
In this section, a toy-model is proposed to show up that a domain wall can behave as a rocket that loses its mass when launched into space.
We simplify Eq.~\ref{2coordq} in the hypothesis of constant $M_{s}$, $K$ and $A_{ex}$: 
\begin{equation}
\label{2coordq3}
    \begin{cases}
        \frac{\alpha J}{\Delta}\dot{q} + J\dot{\phi} +
      H_z^{\mathrm{eff}} M_{s} = 0,\\[2ex]
        -\frac{J}{\Delta}\dot{q} + \alpha J\dot{\phi} + \mathcal{K}_y(t)\sin{\left(2\phi\right)} = 0.
    \end{cases}
\end{equation}
From the second equation of system~\ref{2coordq3}, in the hypothesis of small $\phi$, we derive:
\begin{equation}
    \label{MassEff}
\begin{split}
    &\phi = \frac{J}{2\Delta\mathcal{K}_y(t)}\dot{q},\\
    &\dot{\phi} = \frac{J}{2\Delta\mathcal{K}_y(t)}\ddot{q} + \phi\frac{\dot{\mathcal{K}}_y(t)}{\mathcal{K}_y(t)}.
    \end{split}
\end{equation}
By replacing Eq.~\ref{MassEff}, into the first equation of system~\ref{2coordq3}:
\begin{equation}
\label{ferroRocket2}
    m_{\mathrm{f}}(t)\ddot{q} + \dot{m}_{\mathrm{f}}(t)\dot{q} +\frac{\alpha J}{\Delta}\dot{q} + H_{z}^{\mathrm{eff}} M_{s} = 0,
\end{equation}
where $ m_{\mathrm{f}}(t) = \frac{J^2}{2 \Delta \mathcal{K}_y(t)}$. Remarkably, $\dot{m}_{\mathrm{f}}(t)$ enters the equation of motion in the same way as the damping.

\section{Domain wall motion in non-uniform ferrimagnets}
In the following pages, we aim to expand the description of the domain wall motion in ferrimagnets to the case where spatial non-uniformity is induced. As explained in the Main Text, it is possible to engineer a ferrimagnetic racetrack in such a way that the fundamental parameters vary in space. Here, our model considers non-uniform antiferromagnetic exchange interaction $A$, anisotropy $K$, and sublattice magnetization $M_{1(2)}$ under the main assumption that they vary sufficiently smooth in space, without perturbing the domain wall profile. This simplification is then corroborated by micro-magnetic simulation. The spin Berry phase in a two non-uniform antiferromagnetic-coupled sub-lattices ferrimagnet is defined as \cite{kim2017fast, li2020ultrafast, kim2014propulsion, kim2017self}: 
\begin{equation}
\label{Berry1}
    \mathcal{L}_{\mathrm{B}}(\hat{n}_{\mathrm{1}}, \hat{n}_{\mathrm{2}}) = s_{1}(q) \Bar{a}_{1}(\hat{n}_{1})\cdot\Dot{\hat{n}}_{1} + s_{2}(q)\Bar{a}_{2}(\hat{n}_{2})\cdot\Dot{\hat{n}}_{2},
\end{equation}
where \(s_{1(2)}(q) = \frac{M_{1(2)}(q)}{\gamma_{1(2)}}\), with \(M_{1(2)(q)}\) and \(\gamma_{1(2)}\) the magnetization and the gyromagnetic ratio of each sublattice respectively. \(s_{1(2)}(q)\hat{n}_{1(2)}\) denotes the local spin density per unit of volume, while \(\Bar{a}_{1(2)}\) is the Berry connection. As already mentioned before, the space variation of the parameters does not perturb the domain wall ansatz, therefore, they depends on of the domain wall center $q$. This assumptions is then verified aposteriori by numerical and micromagnetic simulation.
By defining \(s_{\mathrm{T}}(q) = s_{1}(q) + s_{2}(q)\), \(\delta_s(q) = s_{1}(q) - s_{2}(q)\), \(\hat{n} = \frac{\hat{n}_{1} - \hat{n}_{2}}{2}\), \(\hat{m} = \hat{n}_{1} + \hat{n}_{2}\) and expanding Eq.~\ref{Berry1} up to the second order in \(\hat{m}\) \cite{kim2017fast, tveten2016intrinsic, li2020ultrafast}:
\begin{equation}
\label{nonUniformBerry}
    \mathcal{L}_{\mathrm{B}}(\hat{n}, \hat{m}) \approx \frac{s_{\mathrm{T}}(q)}{2}\dot{\hat{n}}\cdot\left(\hat{n}\times\hat{m}\right) + \delta_s(q)\bar{a}\left(\hat{n}\right)\cdot\dot{\hat{n}} + \frac{\delta_s(q)}{4}\dot{\hat{m}}\cdot\left(\hat{n}\times\hat{m}\right).
\end{equation}
In Eq.~\ref{nonUniformBerry}, the first term comes from the cancellation of the spin Berry phases of the two sublattices, while the second and the third terms arise from the residual spin Berry phase. The potential-energy density takes the form \(\mathcal{E}(\hat{n},\hat{m}) = \mathcal{E}(\hat{n}) + \frac{\hat{m}^2}{2\chi}\), where \(\chi\) is the magnetic susceptibility.
By varying the Lagrangian with respect to the magnetization \(\hat{m}\), while neglecting the third term in Eq.~\ref{nonUniformBerry}, we get \(\hat{m} = \frac{s_{\mathrm{T}}(q)}{2}\chi\Dot{\hat{n}}\times\hat{n}\) \cite{tveten2016intrinsic, li2020ultrafast}. Substituting \(\hat{m}\) into Eq. \ref{nonUniformBerry}  leads to \cite{kim2017fast, oh2017coherent, kim2017self, hubert1998magnetic}:
\begin{equation}
\label{nonUniformBerryFinal}
    \mathcal{L}_{\mathrm{B}}(\hat{n}) \approx \frac{\rho (q)}{2}\dot{\hat{n}}^{2} + \delta_{s}(q)\bar{a}\left(\hat{n}\right)\cdot\dot{\hat{n}},
\end{equation}
where \(\rho(q) = \frac{s_{T}^{2}(q) d^2}{4 A(q)}\) parametrizes the inertia associated with the dynamics of \(\hat{n}\). \(d\) is the lattice space. 
The potential-energy density of a non-uniform ferrimagnet is \cite{oh2017coherent, li2020ultrafast, shiino2016antiferromagnetic, hubert1998magnetic}: 
\begin{eqnarray}
\label{potentialEnergy}
    &\mathcal{E}(\hat{n}, q) = \mathcal{E}_{\mathrm{A}}(\hat{n},q) + \mathcal{E}_{\mathrm{K_q}}(\hat{n}, q) + \mathcal{E}_{\mathcal{K}_d}(\hat{n}) + \mathcal{E}_{\mathrm{Z}}(\hat{n}),\\
    \mathcal{E}(\hat{n},q) = &A(q)\left(\nabla\hat{n}\right)^2 +K_{\mathrm{u}}(q)\left(1-n_{\mathrm{z}}^{2}\right) + \mathcal{K}_{d}n_{\mathrm{y}}^{2} - M_s(q)\vec{H}\cdot\hat{n},
\end{eqnarray}
where \(M_s(q)\vec{H}\cdot\hat{n}\) is the Zeeman contribution, with \(M_s(q) = M_{1}(q) - M_{2}(q)\) the saturation magnetization, $\mathcal{K}_d$ is the transverse shape anisotropy. The Gilbert damping effect is taken into account by the Rayleigh dissipation functional \cite{kim2014propulsion, pap2, caretta2020relativistic}:
\begin{equation}
\label{rayleigh}
    \mathcal{R}(\hat{n}) = \frac{s_{\alpha}(q)}{2}\dot{\hat{n}}^{2},
\end{equation}
where \(s_{\alpha} \approx \alpha s_{\mathrm{T}}(q)\), with \(\alpha\) the Gilbert damping parameter. We describe the staggered field \(\hat{n}\) in spherical coordinates:
\begin{equation}
\label{n}
    \hat{n} = \hat{n}\left[\theta(x,q,t), \phi(t)\right] = \left(\sin{\left[\theta(x,q,t)\right]}\cos{\left[\phi(t)\right]}, \sin{\left[\theta(x,q,t)\right]}\sin{\left[\phi(t)\right]}, \cos{\left[\theta(x,q,t)\right]} \right),
\end{equation}

where \(\theta\) is the out-of-plane domain wall angle and \(\phi\) is the precession one.
Substituting Eq.~\ref{n} into Eq. \ref{nonUniformBerryFinal}, \ref{potentialEnergy}, \ref{rayleigh}, the Lagrangian and Rayleigh density functions can be expressed in spherical coordinates:
\begin{eqnarray}
    &\mathcal{L}_{\mathrm{B}}(\theta, \phi, \dot{\theta}, \dot{\phi}) = \frac{\rho(q)}{2}\left[\dot{\theta}^{2} + \sin^{2}{\theta}\dot{\phi}^{2}\right] + \delta_s(q)\phi\dot{\theta}\sin{\theta},\\
    &\mathcal{E}(\theta, \phi, \theta^{'}) = A(q){\theta^{'}}^{2} + K_{\mathrm{u}}(q)\sin^{2}{\theta} + \mathcal{K}_d \sin^{2}{\theta}\sin^{2}{\phi} - M_s(q) H_z\cos{\theta}\\
    &\mathcal{R}(\theta, \phi, \dot{\theta}, \dot{\phi}) = \frac{s_{\alpha}(q)}{2}\left(\dot{\theta}^{2} + \sin^{2}{\theta}\dot{\phi}^{2}\right).
\end{eqnarray}
By applying Euler-Lagrange-Rayleigh equation ($\frac{\partial\mathcal{L}}{\partial\theta} - \frac{\partial}{\partial t}\left[\frac{\partial\mathcal{L}}{\partial\dot{\theta}}\right] + \frac{\partial\mathcal{R}}{\partial\dot{\theta}} = 0$), the domain wall ansatz \(\theta\) is derived: 
\begin{equation} 
    \label{ansatz}
    \theta(x,q,t) = 2\arctan{\left[\exp{\left[{\frac{\gamma(x - q)}{\Delta}}\right]}\right]},
\end{equation}
in the hypothesis of steady-state regime (\(\phi(t) = \phi_{\mathrm{st}}\)). In Eq.~\ref{ansatz} \(\Delta\) is the domain wall width. In Eq.~\ref{ansatz}, \(\gamma = \left(1-\left(\frac{v}{v_{g}(q)}\right)^2\right)^{-1/2}\) is the Lorentz-type contraction factor and \(v_{g}(q) = \frac{2 A(q)}{d s_{\mathrm{T}}}\) is the maximum spin wave velocity, which constitutes the limiting speed for the domain wall. 

\subsection{Three collective coordinates model}
We plug-in the ansatz (Eq. \ref{ansatz}) into the Lagrangian and Rayleigh density functions, integrate in space and then apply the Euler-Lagrange-Rayleigh minimization function:
\begin{equation}
\label{min}
\frac{\partial L(\zeta,\dot{\zeta})}{\partial \zeta} - \frac{d}{dt}\left[\frac{\partial L(\zeta, \dot{\zeta})}{\partial\dot{\zeta}}\right] - \frac{\partial R(\dot{\zeta})}{\partial \dot{\zeta}} = 0,  
\end{equation}
where \(\zeta\) generalizes the domain wall coordinates \(q, \phi, \Delta\). In Eq. \ref{min}:
\begin{equation}
    \begin{cases}
    L(\zeta, \dot{\zeta}) = \int_{V} \mathcal{L}(r,\zeta(t),\dot{\zeta}(t))d^{3} r \hspace{20pt} r \in V,\\
    R(\dot{\zeta}) = \int_{V} \mathcal{R}(r,\dot{\zeta}(t))d^{3}r \hfill r \in V,
    \end{cases}
\end{equation}
therefore:
\begin{equation}
\label{systemEuler}
\begin{cases}
    \frac{\partial\left[L_{\mathrm{B}}-U\right]\left(q\right)}{\partial q} - \frac{\partial}{\partial t}\frac{\partial\left[L_{\mathrm{B}}-U\right]\left(\dot{q}\right)}{\partial\dot{q}} - \frac{\partial R}{\partial \dot{q}} = 0,\\
    \frac{\partial\left[L_{\mathrm{B}}-U\right]\left(\phi\right)}{\partial \phi} - \frac{\partial}{\partial t}\frac{\partial\left[L_{\mathrm{B}}-U\right]\left(\dot{\phi}\right)}{\partial\dot{\phi}} - \frac{\partial R}{\partial \dot{\phi}} = 0,\\
    \frac{\partial\left[L_{\mathrm{B}}-U\right]\left(\Delta\right)}{\partial \Delta} - \frac{\partial}{\partial t}\frac{\partial\left[L_{\mathrm{B}}-U\right]\left(\dot{\Delta}\right)}{\partial\dot{\Delta}} - \frac{\partial R}{\partial \dot{\Delta}} = 0.
    \end{cases}
\end{equation}
Plugging-in the ansatz (Eq.~\ref{ansatz}) into the Euler-Lagrange-Rayleigh functions (Eq.~\ref{systemEuler}), gives: 
\begin{equation}
\frac{\partial\left[L_{\mathrm{B}}-U\right]\left(q\right)}{\partial q} =  \frac{1}{\Delta}\frac{\partial \rho(q)}{\partial q}\dot{q}^2 -2\phi \delta_s'(q)\dot{q} - 2 M_s(q) H_z^{\mathrm{eff}}(q),
\end{equation}

\begin{equation}
-\frac{\partial}{\partial t}\left[\frac{\partial\left[L_{\mathrm{B}}-U\right]\left(\dot{q}\right)}{\partial\dot{q}}\right] =  -2 m_{\mathrm{af}}(q) \ddot{q} + 2\delta_s(q)\dot{\phi} - \frac{2}{\Delta}\frac{\partial \rho(q)}{\partial q}\dot{q}^2 + \frac{2}{\Delta^2}\rho(q)\dot{\Delta}\dot{q} + 2 \phi\delta_s'(q)\dot{q},
\end{equation}

\begin{equation}
    \frac{\partial R}{\partial \dot{q}} = \frac{2\alpha s_{\mathrm{T}}(q)}{\Delta}\dot{q},
\end{equation}

\begin{equation}
    \frac{\partial\left[L_{\mathrm{B}}-U\right]\left(\phi\right)}{\partial \phi} = -2\delta_s (q)\dot{q} - 2\Delta\mathcal{K}_d \sin{2\phi},
\end{equation}

\begin{equation}
    -\frac{\partial}{\partial t}\left[\frac{\partial\left[L_{\mathrm{B}}-U\right]\left(\dot{\phi}\right)}{\partial\dot{\phi}}\right] =  -2\Delta^{2}m_{\mathrm{eff}}(q) \ddot{\phi},
\end{equation}

\begin{equation}
    \frac{\partial R}{\partial \dot{\phi}} = 2\Delta \alpha s_{\mathrm{T}}(q)\dot{\phi},
\end{equation}

\begin{equation}
    \frac{\partial\left[L_{B}-U\right]\left(\Delta\right)}{\partial \Delta} =\left(\dot{\phi}^2 - K_{\mathrm{u}}(q) -\mathcal{K}_d\sin^2{\phi}-\frac{m_{\mathrm{eff}}(q) \dot{q}^2}{\Delta} +\frac{A(q)}{\Delta^2}\right),
\end{equation}

\begin{equation}
    -\frac{\partial}{\partial t}\left[\frac{\partial\left[L_{\mathrm{B}}-U\right]\left(\dot{\Delta}\right)}{\partial\dot{\Delta}}\right] = 0,
\end{equation}

\begin{equation}
    \frac{\partial R}{\partial \dot{\Delta}} = 0.
\end{equation}

$m_{\mathrm{af}}(q) = \frac{\rho (q)}{\Delta}$ is the effective mass of the domain wall, while $H_z^{\mathrm{eff}}(q)$ is the effective driving Zeeman field:
\begin{equation}
  H_z^{\mathrm{eff}}(q) =  \frac{A'(q)}{M_s(q) \Delta} +\frac{\Delta K'_{\mathrm{u}}(q)}{M_s(q)}.
\end{equation}
From the third collective coordinate $\Delta$ we get: 
\begin{equation}
\label{thirdCoord}
    \dot{\phi}^2 - K_{\mathrm{u}}(q) - \mathcal{K}_d\sin^2{\phi} -\frac{m_{\mathrm{af}}(q) \dot{q}^2}{\Delta} +\frac{A(q)}{\Delta^2} = 0.
\end{equation}
\begin{figure}[h]
    \centering
    \includegraphics[width=0.45\linewidth]{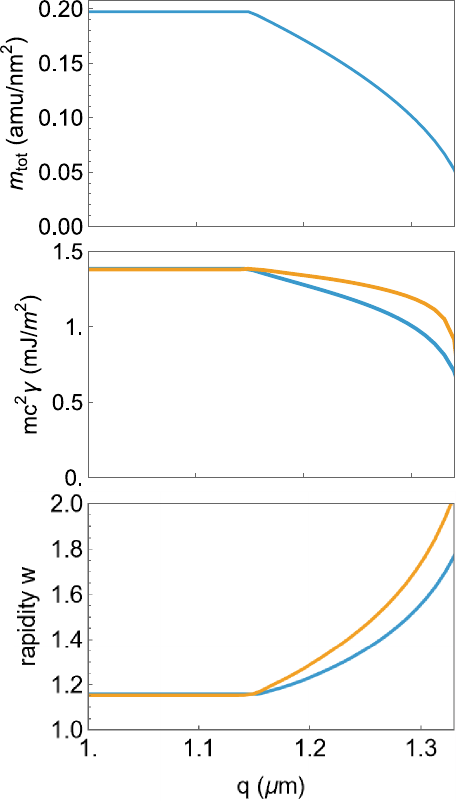}
    \caption{Simulated position dependence of the domain wall mass, kinetic energy and rapidity. Blue line represents the results of simulation without the rocket and magnon velocity gradient terms, while the yellow line simulation includes all mechanisms.}
    \label{fig:placeholder}
\end{figure}
The relativistic domain wall width contraction is derived from the third equation of Eq.~\ref{thirdCoord}:
\begin{equation}
\label{widthContr}
    \Delta(q) = \Delta_0(q) \gamma(q)^{-1},
\end{equation}
where $\gamma(q) = 1/\sqrt{1 - \left(\frac{\dot{q}}{v_{g}(q)}\right)^2}$, $\Delta_0(q) = \sqrt{A(q)/K(q)}$, $v_{g}(q) = \sqrt{\frac{2A (q)}{\rho_{\mathrm{f}}(q) + \rho (q)}}$ is the magnon speed. $\rho_{\mathrm{f}}(q) = \frac{\delta^2_s(q)}{\mathcal{K}_d}$ constitutes the ferromagnetic inertia. 
Consequently, Eq.~\ref{systemEuler} becomes:
\begin{equation}
\begin{cases}
\label{3coord}
m_{\mathrm{af}}(q)\Delta(q)\ddot{\phi} + \frac{\delta_s(q)}{\Delta(q)}\dot{q} + \alpha s_{\mathrm{T}}(q)\dot{\phi} + \mathcal{K}_d \sin{2\phi} = 0,\\
 m_{\mathrm{af}}(q)\ddot{q} - \delta_s(q)\dot{\phi} +\frac{\alpha s_{\mathrm{T}}(q)}{\Delta(q)} \dot{q} + \frac{\partial m_{\mathrm{af}}(q)}{\partial q}\dot{q}^2 -\frac{1}{2\Delta(q)}\frac{\partial \rho(q)}{\partial q}\dot{q}^2 + M_s(q)H_z^{\mathrm{eff}}(q) = 0,\\
     
\end{cases}
\end{equation}

For small angle $\phi$, and negligible $\dot{\phi}$ and $\ddot{\phi}$, the first equation of system~\ref{3coord} simplifies: 

\begin{equation}
\label{phi}
    \phi \approx - \frac{\delta_s(q)}{2\mathcal{K}_d\Delta (q)}\dot{q},
\end{equation}

from which:

\begin{equation}
\label{phiDot}
    \dot{\phi} \approx \left(-\frac{\delta'_{\mathrm{s}}(q)}{\delta_s(q)} + \frac{\Delta'(q)}{\Delta(q)}\right)\frac{\delta_s(q)}{2\mathcal{K}_d\Delta(q)}\dot{q}^2 - \frac{\delta_s(q)}{2\mathcal{K}_d\Delta(q)}\ddot{q}.
\end{equation}

By replacing Eq.~\ref{phi} and Eq.~\ref{phiDot} into the second equation of system~\ref{3coord} and neglecting $\dot{\phi}^2$:
\begin{equation}
\label{eqMotion1}
m_{\mathrm{tot}}(q)\ddot{q} + \frac{\alpha s_{\mathrm{T}}(q)}{\Delta (q)}\dot{q} +  \left[\frac{\partial m_{\mathrm{tot}}(q)}{\partial q} - \frac{m_{\mathrm{f}}(q)\delta_s'(q)}{\delta_s(q)} - \frac{m_{\mathrm{af}}(q)\rho'(q)}{2\rho(q)}\right]\dot{q}^2 + M_s(q)H_z^{\mathrm{eff}}(q) = 0,
\end{equation}
where $m_{\mathrm{tot}}(q) = m_{\mathrm{af}}(q) + m_{\mathrm{f}}(q)$, $m_{\mathrm{af}}(q) = \rho (q)/\Delta (q)$ is the antiferromagnetic mass of the domain wall, $m_{\mathrm{f}}(q) = \delta^2_s(q)/\left[2\mathcal{K}_d\Delta(q)\right]$ is the ferromagnetic mass.




\subsection{Non-uniform exchange interaction and anisotropy}
Here, we assume that anisotropy or antiferromagnetic exchange interaction are varied in space. They act both as a driving field and as the ingredient that induces domain wall mass variation. 
Therefore, Eq.~\ref{eqMotion1} reduces to: 

\begin{equation}
\label{finalEq}
m_{\mathrm{tot}}(q)\ddot{q} = - \frac{\alpha s_{\mathrm{T}}}{\Delta_0 (q)\gamma^2(q)}\dot{q} - \frac{m'_{\mathrm{tot}}(q)}{\gamma^2(q)}\dot{q}^2 -\frac{v'_{g}(q)m_{\mathrm{tot}}(q)\left(1-\frac{2\dot{q}^2}{v^2_{g}(q)}\right)}{v_{g}(q)}\dot{q}^2 -\frac{H_z^{\mathrm{eff}}(q)M_s (q)}{\gamma^3(q)},
\end{equation}
where $\Delta_0(q) = \sqrt{A(q)/K(q)}$ is the non-contracted domain wall width, and $H_z^{\mathrm{eff}}(q) = -2 H_z M_s +A^{'}(q)/\left[\Delta(q)M_s(q)\right] + \Delta (q) K^{'}(q)/M_s(q) $ is the effective magnetic field.

\begin{table*}[t]
\centering
 \caption{Parameters of simulations, illustrated in Fig.~4 of the main text. $[x,y]$ denotes a uniform gradient between $x$ and $y$.}
    \begin{tabular}{@{}lll@{}}
    \hline
    Parameters & $\mathrm{GdFeCo}$ (Fig.~4(a)) &  $\mathrm{GdFeCo}$ (Fig.~4(b))\\
    \hline\hline
    $A$  & $\mathrm{2 \, pJ/m}$ & $[1,4] \, \mathrm{pJ/m}$\\
    \hline
    $M_{\mathrm{A}}$ & $\mathrm{5\cdot 10^{5} \,A/m}$ & $\mathrm{5\cdot 10^{5} \,A/m}$\\
     \hline
    $M_{\mathrm{B}}$ & $\mathrm{4.5\cdot 10^{5} \,A/m}$ & $\mathrm{4.5\cdot 10^{5} \,A/m}$\\
    \hline
    $\alpha$ & $0.004$ & $0.004$\\
    \hline
    $H_z$ & $100 \,  \mathrm{mT}$ & $150 \, \mathrm{mT}$ \\
    \hline
    $K$ & $[14,5]\cdot 10^3\,\mathrm{J/m}^3$ & $10\cdot 10^3\,\mathrm{J/m}^3$\\
     \hline
    $\mathcal{K}_d$ & $40\cdot 10^3\,\mathrm{J/m}^3$ & $40\cdot 10^3\,\mathrm{J/m}^3$\\
    \hline
    d & $\mathrm{0.5\, nm}$ & $\mathrm{0.5\, nm}$ \\
    \hline
    \end{tabular}
    \label{param1}
\end{table*}


\end{document}